\documentclass[11pt,preprint]{aastex}
\usepackage{graphics}

\shorttitle{Variability of Absorption in SDSS J1259+1213}
\shortauthors{Shi et al.}

\begin{document}

\title{Broad Balmer Absorption Line Variability: Evidence of Gas Transverse Motion in the QSO SDSS J125942.80+121312.6}

\author{Xiheng Shi\altaffilmark{1,2}, Hongyan Zhou\altaffilmark{1,3}, Xinwen Shu\altaffilmark{4}, Shaohua Zhang\altaffilmark{1}, Tuo Ji\altaffilmark{1}, Xiang Pan\altaffilmark{2,1}, Luming Sun\altaffilmark{2,1}, Wen Zhao\altaffilmark{2}, Lei Hao\altaffilmark{5}}

\altaffiltext{1}{Polar Research Institute of China, Jinqiao Road 451, Shanghai 200136, China}
\altaffiltext{2}{National Astronomical Observatories, Chinese Academy of Sciences, Beijing 100012, China}
\altaffiltext{3}{Key Laboratory for Research in Galaxies and Cosmology, The University of Science and Technology of China, Chinese Academy of Sciences, Hefei, Anhui, 230026, China}
\altaffiltext{4}{Department of Physics, Anhui Normal University, Wuhu, Anhui, 241000, China}
\altaffiltext{5}{Shanghai Astronomical Observatory, Chinese Academy of Sciences, Shanghai 200030, China}

\begin{abstract}

We report on the discovery of broad Balmer absorption lines variability in the QSO SDSS J125942.80+121312.6, based on the optical and near-infrared spectra taken from the SDSS-I, SDSS-III Baryon Oscillation Spectroscopic Survey (BOSS), and TripleSpec observations over a timescale of 5.8 years in the QSO's rest-frame. The blueshifted absorption profile of H$\beta$ shows a variation of more than 5$\sigma$ at a high velocity portion ($>3000\ \mathrm{km\ s}^{-1}$) of the trough. We perform a detailed analysis for the physical conditions of the absorber using Balmer lines as well as metastable He I and optical Fe II absorptions ($\lambda 4233$ from b$^4$P$_{5/2}$ level and $\lambda 5169$ from a$^6$S$_{5/2}$) at the same velocity. These Fe II lines are identified in the QSO spectra for the first time. According to the photoionization simulations, we estimate a gas density of $n(\mathrm{H})\approx 10^{9.1}\ \mathrm{cm}^{-3}$ and a column density of $N_{\mathrm{col}}(\mathrm{H})\approx 10^{23}\ \mathrm{cm}^{-2}$ for the BOSS data, but the model fails to predict the variations of ionic column densities between the SDSS and BOSS observations if changes in ionizing flux are assumed. We thus propose transverse motion of the absorbing gas being the cause of the observed broad Balmer absorption line variability. In fact, we find that the changes in covering factors of the absorber can well-reproduce all of the observed variations. The absorber is estimated $\sim 0.94$ pc away from the central engine, which is where the outflow likely experiences deceleration due to the collision with the surrounding medium. This scheme is consistent with the argument that LoBAL QSOs may represent the transition from obscured star-forming galaxies to classic QSOs.

\end{abstract}


\section{Introduction}

Broad absorption lines (BALs) are absorptions exhibited in the spectra of QSOs with velocity widths more than $2000\ \mathrm{km\ s}^{-1}$. Their troughs are generally found blueshifted relative to the systemic redshift of QSOs, and believed to originate from the outflow from the inner region of active galactic nuclei (AGNs) with outward velocity in the range several thousand km s$^{-1}$ to 0.2$c$ (Weymann et al. 1981). Thus BALs provide unique diagnostics to the dynamics and physical conditions of the outflowing material. As one of the major forms of AGN feedback, these massive winds may play a critical role in the growth of super massive black holes (SMBHs) and the evolution of their host galaxies (Granato et al. 2004; Scannapieco \& Oh 2004; Hopkins et al. 2008).

It is now well known that BAL systems show both short-term (several months in the QSOs' rest-frame) and long-term (a few years) variability in absorption profiles (Barlow et al. 1992). There are a couple of works in the literature on the variability of high-ionization BALs (HiBALs) (Lundgren et al. 2007: C IV; Gibson et al. 2008, 2010: C IV and Si IV; Capellupo et al. 2011, 2012, 2013: C IV and Si IV; Filiz Ak et al. 2012, 2013: C IV and Si IV). The detection fraction of long-term variations in HiBAL profiles is greater than 50\% (Capellupo et al. 2011; Filiz Ak et al. 2013). The variations are much more frequently found as changes in absorption depth than overall velocity shift. In addition, BAL variability is found to be larger at higher velocity portions of the troughs (Lundgren et al. 2007; Gibson et al. 2008; Capellupo et al. 2011). Two possibilities causing BAL variability are suggested, including the movement of the absorbing medium across our line of sight (LOS) and changes in ionization. The former scenario of varying covering factors is more favored since Barlow et al. (1992), Lundgren et al. (2007), and Gibson et al. (2008) found no strong correlation between BAL variability and continuum changes. However, they could only look at the near-UV variability from the same spectra presenting BAL variations. Without simultaneous observations of extreme-UV flux, the variability of ionizing radiation remained unclear. On the other hand, Wang et al. (2015) claimed a significant negative correlation between the equivalent widths (EWs) of BALs and continuum flux. Furthermore, they found synchronized variations of emission and absorption line EWs. Taking into account the intrinsic Baldwin effect, they believed that BAL variability is driven mainly by changes in ionization in response to continuum variations.

In principle, the changes in the physical conditions of the absorbers at different epochs could definitively reveal the origin of the BAL variability. Unfortunately, HiBAL QSOs, which make up $\sim 80\%$ of all BAL QSOs, are not very useful in studying the physical states of the absorbers. Usually only a single trough of blended doublets is able to be detected. Since the resonant transitions of abundant ions are easily to be saturated, it is very difficult to determine the true optical depths. In cases of partial covering, the observed depths of the HiBAL troughs indicate the covering factors rather than the optical depths. Sub-populations of BALs, such as iron low-ionization BALs (Fe II and Fe III, FeLoBALs) or even rarer hydrogen Balmer series or metastable He I lines, are proved very powerful in constraining the physical conditions of absorbing gas (Wampler et al. 1995; de Kool et al. 2001, 2002; Leighly et al. 2011; Ji et al. 2012; Liu et al. 2015). Vivek et al. (2012, 2014) reported lower occurrence fraction of varying LoBALs and FeLoBALs compared with HiBALs. Zhang et al. (2015) presented a small sample of 28 BAL QSOs, and found that all strong variations of absorption are related with overlapping UV Fe II troughs. The results of photoionization simulations suggest that the absorbers are of high density ($> 10^6\ \mathrm{cm}^{-3}$) and close to the central engines (less than several tens of parsecs). They concluded that these strong variations are due to the movement of absorbing gas across the LOS.

The overlapping UV Fe II absorptions have a major drawback that the blending of numerous lines makes it impossible to measure the ionic column densities on any individual levels, limiting the accuracy of the inferred physical conditions of absorbers. Instead, the well-separated hydrogen Balmer lines and/or metastable He I lines can give a more reliable estimate. In this work, we report on the variability of BALs in the H I Balmer series in the QSO SDSS J125942.80+121312.6 (hereafter as J1259+1213). This object also shows overlapping UV Fe II troughs. For the first time, we identify excited optical Fe II absorption lines in the QSO spectra, which allow for an accurate measurement for Fe$^+$. Using these absorptions, the dynamic structure and the physical conditions of the absorber can be well-defined. We demonstrate that only the scenario of the absorbing medium moving across the LOS can explain the observed BAL variations. This paper is organized as follows. In \S\ref{Observation} we describe the observational features in the SDSS-I, SDSS-III Baryon Oscillation Spectroscopic Survey (BOSS), and TripleSpec spectra, and the identifications of optical Fe II lines. In \S\ref{Measurement} we measure the absorption lines in the BOSS spectrum and estimate the physical properties of the absorber using photoionization simulations. The origin of the profile variations among the SDSS, BOSS and TripleSpec spectra is analyzed in \S\ref{Variation}. In \S\ref{Discussion} we estimate the properties of the QSO and the outflow. A brief summary of the results is presented in \S\ref{Summary}. Throughout this paper we assume a cosmology with $H_{0}=70\mathrm{km\ s}^{-1}\mathrm{Mpc}^{-1}$, $\Omega_{\mathrm{M}}=0.3$ and $\Omega_{\mathrm{\Lambda}}=0.7$.

\section{The Observations and BAL Variability}\label{Observation}

\begin{figure*}
\includegraphics[width=\textwidth]{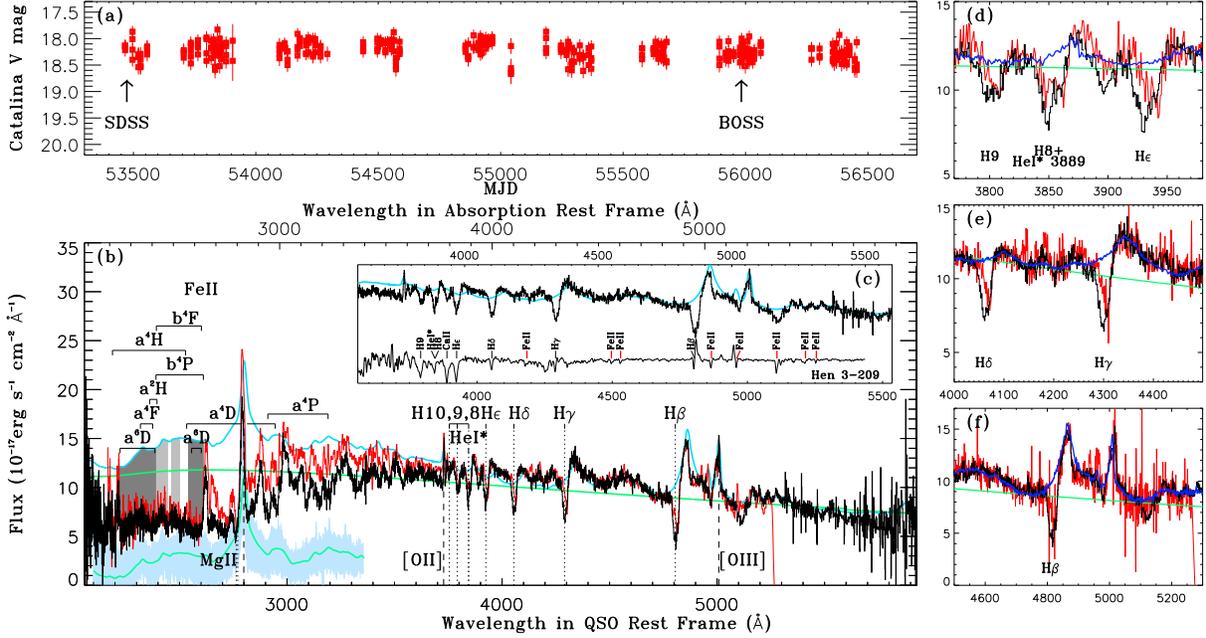}
\caption{Panel (a): the light curve for the Catalina survey. The V-band magnitude shows a variation of $\sim 0.07$ mag between the SDSS and BOSS observations. Panel (b): the BOSS (black line) and smoothed SDSS (red line) spectra of J1259+1213 corrected for Galactic extinction. The cyan line is the SDSS QSO composite spectrum (Vanden Berk et al. 2001) that can best-fit the observation longward of 3770 \AA in the QSO's rest-frame, reddened with SMC-type extinction curve (Gordon et al. 2003) with $E(B-V)=0.187$, and the light green line shows the reddened power-law continuum. The lowest eight terms of Fe$^+$ accounting for the overlapping UV Fe II troughs are labeled, with the wavelengths dominated by the ground term (dark gray) and terms higher than b$^4$P (light gray) being emphasized by shadowed areas. The dark green line below is the UV emission of the SDSS composite spectrum, while the light blue area shows the range of UV emission of the spectra used to construct the unabsorbed optical template. The absorption features detected at rest-frame $\sim 4233$, 4924, 5018 and 5169 \AA are identified as Fe II lines from highly excited levels according to the absorption system associated with stellar wind in emission-line star Hen 3-209 by Naz\'{e} et al. (2006) (detailed in Panel (c)). Panels (d)-(f): the detailed views for variability in Balmer absorption lines from the SDSS to BOSS observations. The blue line represents the unabsorbed optical template using the pair-matching method.\label{SED}}
\end{figure*}

J1259+1213 is a $z=0.751$ QSO with strong broad H I Balmer absorption lines reported by Hall (2007) according to its SDSS DR5 spectrum. The SDSS spectrum, which is corrected for the Galactic extinction using the dust maps by Schlegel et al. (1998) and the mean extinction curve of the Milky Way by Fitzpatrick \& Massa (2007), is shown in Figure \ref{SED} (red line in panel (b)). The deepest Balmer absorption is $\sim 2960\ \mathrm{km\ s}^{-1}$ blueshifted to the systemic redshift and the troughs have a velocity width more than $2300\ \mathrm{km\ s}^{-1}$. The flux blueward of 2600 \AA in the QSO's rest-frame drops due to the absorption of overlapping UV Fe II troughs. Comparing the SDSS spectrum with the reddened SDSS QSO composite spectrum (Vanden Berk et al. 2001) which best-fits the observation in rest-frame optical (longer than 3800 \AA, see Figure \ref{SED} panel (b) cyan line), we find that the relative absorption depths are about 0.5 throughout the troughs between rest-frame 2210 and 2610 \AA. Since the ionization potentials of low-ionization metal ions Mg$^+$ and Fe$^+$ are close to that of atomic hydrogen, we suppose that Mg II and Fe II absorptions have the similar profile to Balmer lines. Thus the absorptions in the ranges of rest-frame 2543.19-2609.56 \AA and 2212.05-2393.38 \AA are dominated by Fe II multiplets UV1 and UV2,3 from the ground term, respectively (the dark gray areas in Figure \ref{SED} panel (b)), while a considerable fraction of absorptions in between should be ascribed to lines from terms with exciting energy $E_{\mathrm{ex}}>2.5\ \mathrm{eV}$ (the light gray areas in Figure \ref{SED} panel (b)). The relative depths of $\sim 0.5$ for these highly excited lines indicate a very high column density for Fe$^+$, for which the ground multiplets should be saturated. However, the spectrum at these wavelengths is non-black, suggesting that the absorber does not fully obscure the background source.

The object was also observed in the SDSS-III BOSS on 2012 February 26, 3.9 years after the SDSS spectroscopic observation in the QSO's rest-frame. The new BOSS spectrum has a wider wavelength coverage and a higher signal-to-noise ratio (S/N). The Galactic extinction-corrected BOSS spectrum is plotted in Figure \ref{SED} panel (b) in black. H I Balmer series from H$\beta$ to H10 are clearly detected.

Comparing the SDSS and BOSS spectra, we find no change in the unabsorbed QSO continuum and emission longward rest-frame 3750 \AA. However, there are clear variations in the absorption profiles of Balmer lines despite the low S/N of the SDSS data. To measure the variations, we use a pair-matching method to define the unabsorbed level for the rest-frame optical spectrum which is better than the rough estimate using the SDSS QSO composite spectrum. For each individual BOSS spectrum from a library of non-BAL QSOs, we fit it with the spectral features of J1259+1213 surrounding optical BALs. If the reduced $\chi ^2 < 1.5$, we consider the fit acceptable, and thus 105 spectra are selected in total. The mean spectrum of these selected spectra will be used as unabsorbed optical template for J1259+1213, and the variance is used to estimate template uncertainty. The method is described in detail in Zhang et al. (2014) and Liu et al. (2015). The resultant template is plotted as blue line in Figure \ref{SED} panels (d)-(f). According to this template, we find the rest-frame EW of H$\beta$ increases from $12.2\pm 1.2$ \AA to $20.3\pm 0.9$ \AA. The error represents 1$\sigma$ uncertainty, including statistical fluctuation in flux and template uncertainty, and the variation is thus greater than 5$\sigma$. In the velocity space as shown in Figure \ref{prof_var} panels (a) and (b), we find that the H$\beta$ profile becomes deeper by 50-100\% blueward $-3000\ \mathrm{km\ s}^{-1}$. That the absorption profile changes more significantly at higher blueshifting velocity is consistent with the results observed for high-ionization lines.

With the improved data quality and unabsorbed template, we identify that He I* $\lambda 3889$ also contributes to the H8 absorption trough at rest-frame 3850 \AA. In the BOSS spectrum, the trough has an EW of $5.52\pm 0.21$ \AA, larger than H$\epsilon$ ($4.96\pm 0.19$ \AA) which should be $\sim 60\%$ stronger than H8.

A followup TripleSpec near-infrared (NIR) spectroscopic observation was performed on 2015 May 26 (see Figure \ref{TPSP} top panel), 1.8 years after the BOSS observation in the QSO's rest-frame. The spectrum covers from rest-frame 5551 to 14065 \AA. The broad absorption trough of H$\alpha$ is also detected at the same radial velocity as higher-order Balmer lines in the SDSS and BOSS spectra. At high blueshifted velocity portion of the trough, nearly all background flux is absorbed, indicating a further variation of Balmer lines along with the time.

J1259+1213 was also monitored photometrically by the Catalina survey from 2005 May to 2013 July. The light curve is plotted in Figure \ref{SED} panel (a). The difference of V-band magnitude between the SDSS and BOSS observations is $0.07\pm 0.17$. The filter of Catalina V-band is centered at $\sim 3100$ \AA in the QSO's frame. We find the V-band flux decreases from the SDSS to the BOSS spectra by 18\%, corresponding to 0.18 mag. This result is consistent with the photometric data within errors, indicating no severe problem existing in the flux calibration for the spectroscopic data. Thus, we conclude that the near-UV to optical continuum shows no significant variability. However, because of lacking simultaneous observations of extreme-UV flux, we know little about the ionizing radiation variability. The changes of ionization as a cause of BAL variations cannot be definitively ruled out. To reveal the origin of BAL variations, further investigations of the physical conditions of the absorber are required.

\section{Measurement and Modeling the Absorber in BOSS observation}\label{Measurement}

To explore the nature of BAL variability, we start with the BOSS observation, which has much higher S/N. Broad hydrogen Balmer absorption lines from H$\beta$ to H10 are well-detected in the BOSS spectrum, offering a unique opportunity to determine the velocity structure of H I absorber and column density on the hydrogen $n=2$ shell.

In the spectrum we find the residual flux of the saturated UV Fe II troughs is larger than the UV Fe II emission bump in the SDSS QSO composite spectrum (the underlying dark green line in Figure \ref{SED} panel (b)) by 60-100\%. Furthermore, among the 105 non-BAL QSOs selected resembling the optical spectrum of J1259+1213, 44 cover the UV Fe II bump without narrow absorptions. The light blue area in Figure \ref{SED} panel (b) shows the UV emission lines of these spectra. The observed spectrum also lies above all these features. The excess could come from unobscured accretion disk. This suggests that the absorber only covers part of the continuum source and little of the broad emission-line region (BELR), as the latter is two orders of magnitude larger in size.

Removing the emission lines of optical template from observed flux and normalizing the rest by power-law continuum, we show the profiles of H$\beta$, H$\gamma$, H$\delta$, and H10 in Figure \ref{prof_var} panel (a). The normalized intensity is

\begin{equation}
I(v)=[1-C_f(v)]+C_f(v)e^{-\tau (v)},\label{eq1}\\
\end{equation}

where $C_f(v)$ is the covering factor and $\tau$ is true optical depth as functions of radial velocity. For transitions from the same level of the given ion, the values of $\tau (v)$ are proportional to $gf\lambda$, where $g$ is the statistical weight, $f$ is the oscillator strength, and $\lambda$ is the rest wavelength of the transition.

Applying Eq.\ref{eq1}
to H$\beta$, H$\gamma$, H$\delta$ and H10, we can derive $C_f(v)$ and $\tau (v)$ through the Balmer troughs. The results are shown in Figure \ref{prof_var} panels (c) and (d). The integrated column density on the hydrogen $n=2$ shell through the profile is $7.01\pm 0.54\times 10^{15}\ \mathrm{cm}^{-2}$. By removing the contribution of the predicted H8 absorption from the trough at rest-frame 3850 \AA and assuming $C_f(v)$ and fractional distribution $\tau (v)/\int \tau(v)\,dv$ extracted from Balmer lines are also valid for He I* $\lambda 3889$, we derived that the column density on metastable He$^0$ 2$^3$S is $1.28\pm 0.09\times 10^{15}\ \mathrm{cm}^{-2}$.

Ji et al. (2015) presented a detailed investigation on the He I* ionization structure using photoionization simulations. They found that if the irradiated medium is thick enough and the ionizing front is well-developed, the strength of ionizing flux (represented by ionization parameter $U$) at the illuminated surface can be solely determined by the column density of He I*. From the Figure 10 in Ji et al. (2015), we obtain $\log U=-1.40\pm 0.05$ for the BOSS spectrum.

To further constrain the physical properties of the absorber, other absorption lines in addition to Balmer and He I* are required. A couple of weaker absorption features are detected in the BOSS spectrum between the rest-frame 4200 and 5300 \AA (see Figure 1, inset panel (c)), though not seen in the SDSS spectrum due to the low S/N. The most credible ones are troughs at 4200, 4880, 4970, and 5120 \AA, detected at a statistical significance greater than $3\sigma$. The strongest one is at 5120 \AA, having a profile with $\mathrm{FWHM}>1500\ \mathrm{km\ s}^{-1}$, which is comparable to the Balmer absorption lines and suggests that it could be part of the intrinsic blueshifted system. If these lines are associated with Balmer transitions, the rest-frame wavelengths are at 4233, 4924, 5018, and 5169 \AA respectively.

In fact, absorptions at these wavelengths have never been reported in the QSO spectra in previous works. However, in other cases like extremely strong stellar wind, similar features have been identified before. In the spectrum of emission-line star Hen 3-209, Naz\'{e} et al. (2006) identified a blueshifted Fe II absorption system associated with H I and He I, consisting of lines at rest wavelengths 4233, 4549, 4584, 4924, 5018, 5169, 5276, and 5317 \AA (see the inset panel (c) in Figure \ref{SED}). These absorptions are also present in J1259+1213 but are too weak allowing for meaningful measurements except lines at 4233 and 5169 \AA. According to the atomic data from NIST{\footnote{http://www.nist.gov/pml/data/asd.cfm}}, we identify these lines as arising from the excited levels of Fe$^+$, with $E_{\mathrm{ex}}$ varying from 2.58 to $3.20\ \mathrm{eV}$. The details are listed in Table \ref{OptFeII}.

These isolated optical Fe II lines are useful to better constrain the physical state of Fe$^+$, compared with the overlapping UV Fe II troughs. Assuming Fe II* $\lambda 4233$ and $\lambda 5169$ have the same profile as Balmer lines, we estimate that column densities are $4.31\pm 4.06\times 10^{15}\ \mathrm{cm}^{-2}$ and $5.13\pm 1.29\times 10^{15}\ \mathrm{cm}^{-2}$ for Fe$^+$ b$^4$P$_{5/2}$ and a$^6$S$_{5/2}$ levels, respectively.

We use the photoionization code CLOUDY (version 10.00, last described by Ferland et al. 1998) to simulate the ionization process in the absorbing medium, assuming a slab-shaped geometry, unique density, and homogeneous chemical composition of solar values. The incident spectral energy distribution (SED) applied is a combination of a UV bump described as $\nu^{\alpha_{\mathrm{UV}}}\mathrm{exp}(-h\nu /kT_{\mathrm{BB}})\mathrm{exp}(-kT_{\mathrm{IR}}/h\nu)$ and power-law $a\nu^{\alpha_{\mathrm{X}}}$, incorporated in CLOUDY. This is considered typical for observed AGN continuum. The UV bump is parameterized by UV power-law index $\alpha_{\mathrm{UV}}=-0.5$, and exponentially cut off with temperature $T_{\mathrm{BB}}=1.5\times 10^5\ \mathrm{K}$ at high energy and $kT_{\mathrm{IR}}=0.01\mathrm{Ryd}$ at infrared. The power-law component has an index $\alpha_{\mathrm{X}}=-2$ beyond 100 keV, and $-1$ between 1.36 eV and 100 keV. The overall flux ratio of X-ray to optical is $\alpha_{\mathrm{OX}}=-1.4$.

The resultant column densities on the hydrogen $n=2$ shell and Fe$^+$ excited levels as functions of $n(\mathrm{H})$ and $N_{\mathrm{col}}(\mathrm{H})$ for ionizing flux of $\log U=-1.4$ are plotted in Figure \ref{simu}, where the measured values and uncertainties are presented in the colored areas. Our photoionization simulations suggest that model with $n(\mathrm{H})\sim 10^{9.1\pm 0.1}\ \mathrm{cm}^{-3}$ and $N_{\mathrm{col}}(\mathrm{H})\sim 10^{23.0\pm 0.1}\ \mathrm{cm}^{-2}$ is able to reproduce measured column densities of Balmer and optical Fe II lines.

The detailed Fe$^+$ atomic model incorporated in CLOUDY can also predict column densities on various levels contributing to the overlapping UV and optical Fe II absorption troughs between 2200 and 3300 \AA. To check if the model result is consistent with UV Fe II, we need to generate a synthetic spectrum. We assume that all individual absorption lines have the same profile as Balmer lines. This means, for a given Fe II transition, the ionic column density on the lower level predicted by CLOUDY will be distributed to different outflow velocities to evaluate the optical depth $\tau_{\mathrm{FeII}}(v)$, following the fractional column density distribution $\frac{dN_{\mathrm{col}}/dv}{N_{\mathrm{col}}}$ where $dN_{\mathrm{col}}/dv$ is proportional to $\tau (v)$ from Balmer series. Then the effect of partial covering in considered as $f_{\mathrm{model}}=f_{\mathrm{em}}+f_{\mathrm{conti}}\times (1-C_f+C_f e^{-\tau})$ to get the model flux, where $f_{\mathrm{em}}$ and $f_{\mathrm{conti}}$ are unabsorbed fluxes of emission lines and continuum in J1259+1213 respectively. Since little is known about the unabsorbed UV flux of J1259+1213, we have to estimate it. An initial try is to use the best-fitting SDSS QSO composite spectrum shown in Figure \ref{SED} as an unabsorbed level. The resultant synthetic spectrum is plotted in Figure \ref{synth_spec} panel (a). The major features are reproduced, yet minor inconsistency is not negligible. The main cause is that the unabsorbed flux of J1259+1213 naturally differs from the average. A better method is required, like Leighly et al. (2011) deemed the best matched non-BAL QSO spectrum as unabsorbed template to measure He I* $\lambda 10830$ absorption. We employ the spectrum of $z_{\mathrm{em}}=0.810$ QSO SDSS J155635.81+160021.2 from the BOSS catalog as an unabsorbed template, which best reproduces the observation when combined with the absorption model. The uncertainty of the synthetic model spectrum includes the uncertainty introduced from photoionization model parameters $U$, $n(\mathrm{H})$, and $N_{\mathrm{col}}(\mathrm{H})$ along with the uncertainty of the template. The latter is estimated using the variance of the 44 non-BAL spectra which are selected to construct the optical template and extend it to UV Fe II. In Figure \ref{synth_spec} panel (b) we plot the synthetic spectrum and its uncertainty. The model reproduces the observed spectrum better than the simple SDSS composite spectrum, though there are still slight deviations at some wavelengths.

Based on the luminosity and SED of the ionizing continuum, we can derive the distance of the absorbing gas to the central source as $\frac{L(<912)}{4\pi r_{\mathrm{abs}}^2}=Un(\mathrm{H})c\overline{E_{\mathrm{ph}}(<912)}$, where $L(<912)$ is the ionizing luminosity of the continuum source and $\overline{E_{\mathrm{ph}}(<912)}$ is the average energy for all ionizing photons. With the continuum flux at 1215.67 \AA from the best-fitting SDSS composite spectrum presented in Figure \ref{SED} and the incident model SED used in simulations, the extrapolated extinction-corrected monochromatic luminosity of ionizing continuum in J1259+1213 at Lyman limit is $L_{\nu}(912)=9.8\times 10^{29}\ \mathrm{erg s}^{-1}\mathrm{Hz}^{-1}$. Using this value as the unabsorbed continuum luminosity at 1 Ryd, we obtain the $L(<912)$. $\overline{E_{\mathrm{ph}}(<912)}$ is also evaluated according to SED. With $\log U=-1.4$ and $\log n(\mathrm{H})(\mathrm{cm}^{-3})=9.1$ from the best-fitting model, the distance of the inner surface of the absorbing gas is $r_{\mathrm{abs}}=0.94\ \mathrm{pc}$. The uncertainty of the distance introduced from $U$ and $n(\mathrm{H})$ is around 14\%. In addition, the distance also depends on the AGN ionizing luminosity, which is uncertain due to the extrapolation of power-law continuum. A change of 100\% for the luminosity can lead to a change of 41\% for distance, making it possibly the major source of uncertainty.

\section{Variation of absorption from SDSS to BOSS}\label{Variation}

Based on the inferred physical conditions of the absorbing gas from Balmer, He I* and Fe II lines in the BOSS spectrum, we can investigate the various scenarios responsible for the observed BAL variations in detail.

We first check for the possibility of the changes in incident continuum flux, which would affect the ionization structure and alter the ionic column densities. The medium responds to continuum variations in the recombination timescale $(\alpha n_{\mathrm{e}})^{-1}$, where $\alpha\approx 4\times 10^{-13}\ \mathrm{cm}^{-3}\mathrm{s}^{-1}$ is the recombination coefficient of H$^0$. In J1259+1213, the absorber is so dense that the timescale is as short as $\sim 2\times 10^3\ \mathrm{s}$. Thus, the time interval between the SDSS and BOSS observations is long enough to allow for the ionization state of the medium responding to any changes of ionizing flux.

\begin{figure}
\includegraphics[width=0.5\columnwidth]{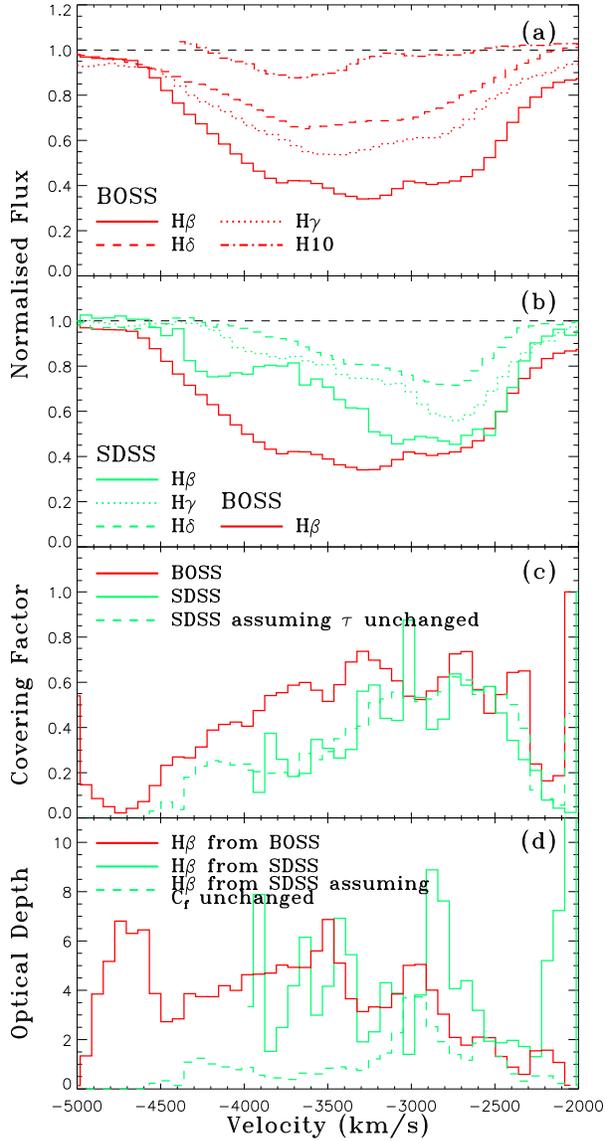}
\caption{Panels (a) and (b): the smoothed and normalized absorption profiles for H$\beta$, H$\gamma$, H$\delta$, et al., in the BOSS and SDSS observations. Since the absorber only covers part of the accretion disk, the profiles are constructed by removing the emission lines of the unabsorbed template and then dividing the rest by power-law continuum. Panel (c): the covering factors as functions of radial velocity derived using Eq.\ref{eq1}. The green dashed line represents the covering factor assuming the true optical depth of H$\beta$ in the SDSS observation is the same as in the BOSS. Panel (d): the optical depths as functions of radial velocity derived using Eq.\ref{eq1}. The green dashed line represents the optical depth, assuming the covering factor of H$\beta$ in the SDSS spectrum remains unchanged.}\label{prof_var}
\end{figure}

However, the variations of ionizing flux should not change the geometry of the absorber (e.g. the covering factor as a function of radial velocity). For H$\beta$ absorption in the SDSS spectrum, if we assume the distribution of covering factor is the same as that we derived from the BOSS data, the optical depth of H$\beta$ at $v_{\mathrm{rad}}<-3000\ \mathrm{km\ s}^{-1}$, $\tau (v)=-\ln \frac{C_f(v)-(1-I(v))}{C_f(v)}$, becomes less than 1 as the apparent depth ($1-I(v)$) decreases (see dashed green line in Figure \ref{prof_var} panel (d)). The measured column density is $N_{\mathrm{col}}(\mathrm{H}^0_{n=2})=2.31\pm 0.37\times 10^{15}\ \mathrm{cm}^{-2}$, and $N_{\mathrm{col}}(\mathrm{He I*})=9.35\pm 0.91\times 10^{14}\ \mathrm{cm}^{-2}$ when the contribution of H8 is removed from the blended trough at rest-frame 3850 \AA in the SDSS spectrum.

The smaller column density of He I* means that the absorber is irradiated by the ionizing flux with $\log U=-1.60\pm 0.05$ during the SDSS observation. If the changes in ionizing flux are assumed, the density and total column density of the medium should remain constant between the SDSS and BOSS observations. Therefore, the physical parameters $n(\mathrm{H})=10^{9.1\pm 0.1}\ \mathrm{cm}^{-3}$ and $N_{\mathrm{col}}(\mathrm{H})=10^{23.0\pm 0.1}\ \mathrm{cm}^{-2}$ derived from the BOSS spectrum should also be able to reproduce SDSS observation. However, under the ionizing flux of $\log U=-1.60\pm 0.05$, the predicted column density is $N_{\mathrm{col}}(\mathrm{H}^0_{n=2})=5.8\pm 1.3\times 10^{15}\ \mathrm{cm}^{-2}$, which is much larger than the measured value of $2.31\pm 0.37\times 10^{14}\ \mathrm{cm}^{-2}$. It is least unlikely that the observed line variations are due to the changes in ionization state of the absorbing gas.

Then we check for the alternative scenario in which the variations in BALs are mainly caused by the movement of the absorber. In this case, the changes of the observed profile between different epochs are due to the changed fraction of ionization source obscured by the medium or its sub-structure.

In Figure \ref{prof_var} panel (c), we plot the covering factor as a function of radial velocity, derived as $C_f(v)=\frac{1-I(v)}{1-e^{-\tau (v)}}$, for H$\beta$ absorption in the SDSS spectrum, assuming the optical depth does not vary. In this case, the decrease of apparent depth directly indicates the decrease of covering factor. On the other hand, we can apply Eq.\ref{eq1} to the multiple Balmer lines in the SDSS spectrum to derive the optical depth and covering factor as we did for the BOSS data. The results are also presented in Figure \ref{prof_var} as green solid lines. Despite the low S/N of the SDSS data, the measured covering factor is nearly the same as that derived when assuming no change in optical depth (Figure \ref{prof_var} panel (c)) which is about a factor of 2 lower than that measured in the BOSS spectrum (red solid line). Furthermore, as shown in Figure \ref{prof_var} panel (d), the measured optical depth, though affected by the noise in the SDSS spectrum, is in good agreement with that measured from the BOSS data. In contrast, the measured optical depth is clearly different from that derived assuming constant covering factor. This result strongly suggests the transverse movement of absorbing gas (and thus the change of covering factor) as the origin for the observed BAL variability in J1259+1213.

\begin{figure}
\includegraphics[width=0.5\columnwidth]{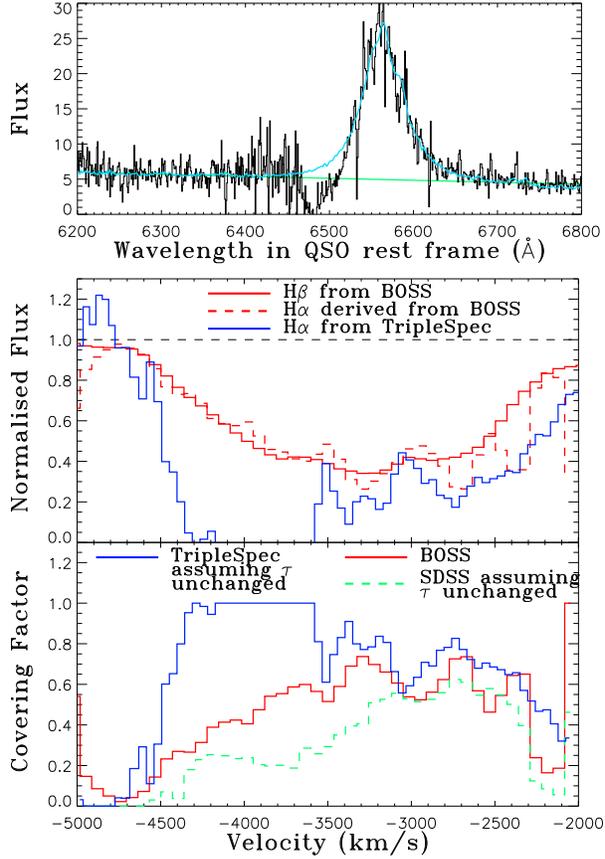}
\caption{Followup TripleSpec NIR observation. Top panel: the observed spectrum around H$\alpha$. The cyan line represents the unabsorbed template and the green line represents the continuum. Middle panel: the normalized profile for H$\alpha$ from the TripleSpec (blue), compared with H$\beta$ from the BOSS (red solid) and H$\alpha$ derived from Balmer series from the BOSS (red dashed line). At around $-4000\ \mathrm{km\ s}^{-1}$, all continuum radiation is absorbed. Bottom panel: the covering factor for H$\alpha$ in the TripleSpec (blue) assuming the optical depth of the absorber does not vary from the BOSS observation, compared with the covering factor of Balmer lines from the BOSS (red solid line), and the covering factor from the SDSS assuming the optical depth of the absorber does not change (green dashed line). The covering factor at high outward velocity ($<-3000\ \mathrm{km\ s}^{-1}$) increases continuously.\label{TPSP}}
\end{figure}

\begin{figure}
\includegraphics[width=0.5\columnwidth]{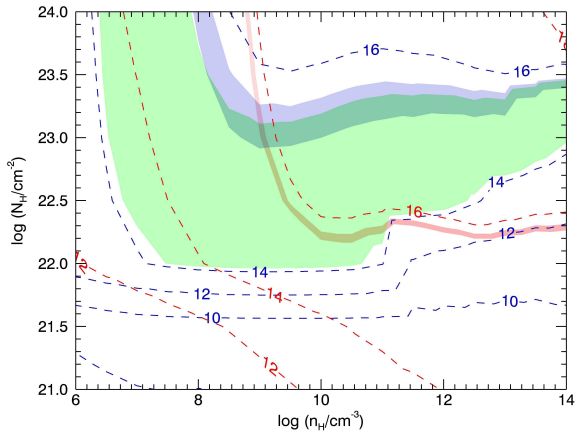}
\caption{Predicted ionic column densities (dashed lines) from the photoionization simulations by CLOUDY with $U=10^{-1.4}$ as functions of $n(\mathrm{H})$ and $N_{\mathrm{col}}(\mathrm{H})$. The attached numbers are logarithms of ionic column densities. The red contours show $\mathrm{H}^0_{n=2}$, and the blue ones show $\mathrm{Fe^+_{a^6S_{5/2}}}$. The colored areas represent measured column densities for $\mathrm{H}^0_{n=2}$ (red), $\mathrm{Fe^+_{b^4P_{5/2}}}$ (green) and $\mathrm{Fe^+_{a^6S_{5/2}}}$ (blue). The model of $U=10^{-1.4}$, $n(\mathrm{H})=10^{9.1}\ \mathrm{cm}^{-1}$ and $N_{\mathrm{col}}(\mathrm{H})=10^{23}\ \mathrm{cm}^{-2}$ can reproduce the measured strengths of Balmer lines, He I* and optical Fe II simultaneously.\label{simu}}
\end{figure}

\begin{figure}
\includegraphics[width=0.8\textwidth]{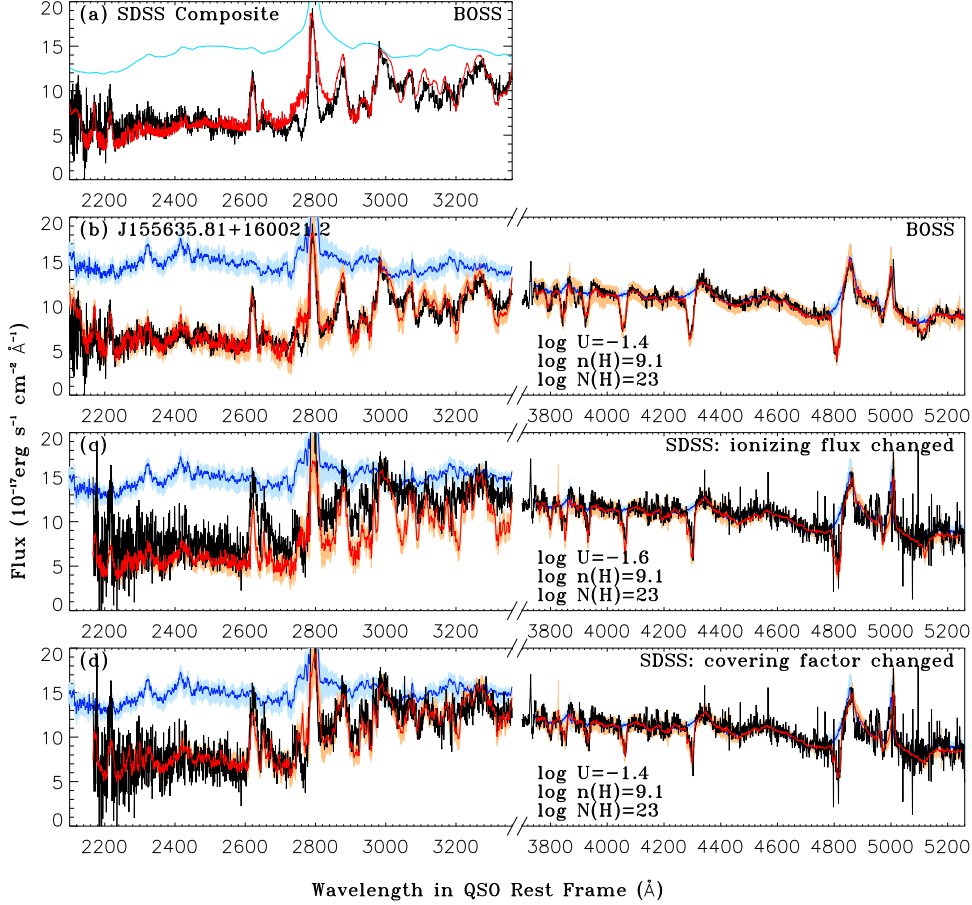}
\caption{Synthetic spectra (red lines) constructed according to the photoionization models to be compared with the observed UV Fe II and Balmer absorptions. Besides H I, He I*, and Fe II, the contributions from Mg I, Mg II, Ti II, Cr II, Mn II, Co II, and Ni II are also included. All transitions are assumed to have the same profile as Balmer lines. Since in rest-frame UV band, little is known about the unabsorbed spectrum of J1259+1213 due to the overlapping UV Fe II troughs, we use the SDSS QSO composite spectrum (cyan line) to estimate the unabsorbed level first (see panel (a)). In the following panels, the spectrum of SDSS J155635.81+160021.2 (blue lines in the left half of all panels) is employed as unabsorbed UV template and reddened using SMC-type extinction law. In panels (a) and (b), the model best-fitting the BOSS observation is plotted. Panel (c) exhibits the synthetic model spectrum assuming that the absorber during the SDSS observation has the same $n(\mathrm{H})$, $N_{\mathrm{col}}(\mathrm{H})$ and covering factor as that during the BOSS while ionizing flux is $U=10^{-1.6}$. In panel (d), we show the model spectrum assuming the absorber at the SDSS observation has the same physical properties ($U$, $n(\mathrm{H})$ and $N_{\mathrm{col}}(\mathrm{H})$) as that in the BOSS while the covering factor at high outward velocity changes. The uncertainty of the template is evaluated using the variance of non-BAL spectra employed to construct optical template, and is plotted in light blue. The uncertainty of synthetic model spectra (in orange) includes the uncertainty of the template and the uncertainty from photoionization simulation parameters $U$, $n(\mathrm{H})$, and $N_{\mathrm{col}}(\mathrm{H})$. It is clear that the scenario of the moving absorber rather than changing ionizing flux can explain the variations.\label{synth_spec}}
\end{figure}

The TripleSpec NIR data also support the picture of transverse movement. In Figure \ref{TPSP} middle panel we plot the normalized H$\alpha$ absorption spectrum. With the knowledge of optical depth and covering factor as functions of radial velocity for H$\beta$ in the BOSS spectrum, we can derive the profile of H$\alpha$ absorption during the BOSS observation. We find almost identical H$\alpha$ profile at $v_{\mathrm{rad}}>-3500\ \mathrm{km\ s}^{-1}$ during the TripleSpec observation. But at higher outward velocity nearly all continuum flux is absorbed. This can not be explained by the growth of column density as a result of changes in ionizing flux. Conversely, if we assume the covering factor at a higher velocity continues to increases between the BOSS and TripleSpec observations (see Figure \ref{TPSP} bottom panel, the absorber tends to obscure the whole continuum source at $v_{\mathrm{rad}}<-3500\ \mathrm{km\ s}^{-1}$), the BAL variability between SDSS-I, SDSS-III BOSS, and TripleSpec data can be easily understood.

Furthermore, we construct two synthetic spectra for UV Fe II absorption, one corresponding to the scenario of changes in ionizing flux, and the other for the changes in covering factor. In Figure \ref{synth_spec} panel (c) we show the synthetic spectrum for the model with $n(\mathrm{H})=10^{9.1}\ \mathrm{cm}^{-3}$, $N_{\mathrm{col}}(\mathrm{H})=10^{23}\ \mathrm{cm}^{-2}$ and unvaried covering factor, assuming only the incident ionizing flux changes to $\log U=-1.6$. It can be seen that the predicted Fe II absorption troughs from 2600 to 3400 \AA are much deeper than the observed ones. In fact, we test and find that in case of changes only in the ionizing flux, no model can reproduce the observed BAL features in the SDSS spectrum. In Figure \ref{synth_spec} panel (d), the synthetic spectrum constructed, assuming a decrease in covering factor at $v<-3000\ \mathrm{km\ s}^{-1}$, can reproduce the SDSS data quite well. Note that except for the covering factor, the model parameters are the same as those for the BOSS data. We thus attribute the BAL variability in J1259+1213 to gaseous medium moving across our LOS, possibly related to a substantial outflowing structure.

\section{Discussion}\label{Discussion}

The mass of central SMBH can be derived using the relation $\log (M_{\mathrm{BH}}/10^6\ M_{\odot})=(1.39\pm 0.14)+0.5\log (L_{5100}/10^{44}\ \mathrm{erg\ s}^{-1})+(1.09\pm 0.23)\log[\mathrm{FWHM}(\mathrm{H}\beta)/1000\ \mathrm{km\ s}^{-1}]$ from Wang et al. (2009). $L_{5100}$ is the monochromatic AGN luminosity at the rest-frame 5100 \AA which is $1.54\pm 0.07\times 10^{45}\ \mathrm{erg\ s}^{-1}$ for J1259+1213, and FWHM(H$\beta$) is the full width at half maximum for broad H$\beta$ emission component. Since in the spectra of J1259+1213 the broad absorption troughs superimpose the blue wing of Balmer peaks, we use the unabsorbed template shown in Figure \ref{SED} (panel (f)) to extract the profile of Balmer lines. The H$\beta$ is modeled using one Gaussian profile for narrow emission and three Gaussians for broad emission. The resulting FWHM for the whole broad H$\beta$ component is $3631\ \mathrm{km\ s}^{-1}$. Therefore, we obtain $M_{\mathrm{BH}}=4.0\pm 1.7\times 10^8\ M_{\odot}$. Adopting the correction factor by Runnoe et al. (2012), the bolometric luminosity is $L_{\mathrm{bol}}=(8.1\pm 0.4)\times L_{5100}=1.25\pm 0.08\times 10^{46}\ \mathrm{erg\ s}^{-1}$. Assuming an accretion efficiency of 0.1, the mass accretion rate is then $\dot{M}_{\mathrm{BH}}\approx 2.2\ M_{\odot}\ \mathrm{yr}^{-1}$.

The radius of the accretion disk that effectively contributes to the continuum radiation at 4863 \AA can be estimated from $\sigma T_{\mathrm{eff}}^{4}=\frac{3GM\dot{M}}{8\pi R^{3}}f(R,a)$ (Eq. 2 in Collin et al. 2002). With $k_{\mathrm{B}}T_{\mathrm{eff}}=h\nu$ and the boundary condition $f(R,a)\approx 1$, we obtain $R(4863)=2.7\times 10^{15}\ \mathrm{cm}$. If the increase of a covering factor at $v_{\mathrm{rad}}\approx -3700\ \mathrm{km\ s}^{-1}$ from 0.2 during the SDSS observation to 0.6 during the BOSS observation stands for a continuous movement of absorber across our LOS, the transverse velocity is $v_{\bot}\approx 1.8\times 10^2\ \mathrm{km\ s}^{-1}$, which is much smaller than the Keplerian velocity $1.4\times 10^3\ \mathrm{km\ s}^{-1}$ at $r_{\mathrm{abs}}$ ($\sim 0.94\ \mathrm{pc}$).

In the model where the outflow is accelerated radially, the conservation of angular momentum is suggested, $v_{\bot}\propto r^{-1}$. Therefore, we find that at a smaller distance, $r_{\mathrm{eq}}\approx 4.8\times 10^{16}\ \mathrm{cm}$, the extrapolated rotational velocity equals the Keplerian velocity which varies as $r^{-0.5}$. Since the largest transverse velocity is detected at an outward velocity of $\sim 3700\ \mathrm{km\ s}^{-1}$, the corresponding absorbing gas may locate around the illuminated surface of the outflow, closer to the center than the gas at a lower outward velocity. Therefore, it seems that the outflow arises from the surface of the inner disk at $r_{\mathrm{eq}}$, being accelerated radially up to $3700\sim 4000\ \mathrm{km\ s}^{-1}$, the largest outward velocity we detect in the absorption troughs, at around 1 pc, and then gradually being decelerated. The distance corresponding to the largest outward velocity is close to the evaporation radius of $R_{\mathrm{evap}}=1.3L_{\mathrm{UV,46}}^{1/2}T_{1500}^{-2.8}\mathrm{pc}\approx 0.77\ \mathrm{pc}$ (Barvainis 1987), where $L_{\mathrm{UV,46}}$ is the UV luminosity in unit of $10^{46}\mathrm{erg\ s}^{-1}$ which is approximated using $\lambda L_{\lambda}(1450)$, and $T_{1500}$ is the grain evaporation temperature in unit of 1500 K which is $\approx 1$. The deceleration could be due to the collision with the surrounding interstellar medium. This picture is consistent with the suggestion that LoBAL QSOs may represent the transition from obscured star-forming galaxies to classic QSOs when the outflow blows away the dusty envelope (Canalizo \& Stockton 2001, Di Matteo et al. 2005).

The mass of an individual outflow element can be estimated as $M_{\mathrm{abs}}=\mu m_{\mathrm{p}}N_{\mathrm{col}}(\mathrm{H})S$, where $\mu$ is the mean atomic mass per proton, $m_{\mathrm{p}}$ is the mass of proton, and $S$ is the projection area of the absorber. Its dynamic timescale can be approximated as $2r_{\mathrm{abs}}/v_{\mathrm{rad,max}}$. Assuming all wind elements have similar physical properties, the total mass outflow rate can be expressed as the sum of all elements at a given time, which is $\sum_i \frac{\mu m_{\mathrm{p}}N_{\mathrm{col},i}(\mathrm{H})S_i v_{\mathrm{rad,max},i}}{2r_{\mathrm{abs,i}}}\approx \frac{\mu m_{\mathrm{p}}N_{\mathrm{col}}(\mathrm{H}) v_{\mathrm{rad,max}}}{2r_{\mathrm{abs}}} \sum_i S_i$. If these wind elements cover a solid angle of $\Omega$ in the sky relative to the continuum source, $\sum_i S_i=r_{\mathrm{abs}}^2\Omega$, then

\begin{equation}
\dot{M}_{\mathrm{wind}}=\frac{1}{2} r_{\mathrm{abs}}\Omega \mu m_{\mathrm{p}}N_{\mathrm{col}}(\mathrm{H})v_{\mathrm{rad,max}}\label{eq3}
\end{equation}

$\Omega$ cannot be derived directly from the observation of J1259+1213. However, we can give a rough estimate for $\Omega$ according to the detection fraction of BAL QSOs, which is suggested $\sim 40\%$ by considering the spectroscopic incompleteness and selection bias (Allen et al. 2011). Thus, we obtain $\dot{M}_{\mathrm{wind}}\approx 10\ M_{\odot}\ \mathrm{yr}^{-1}$. The kinetic energy carried by the wind is $\dot{E}_{\mathrm{k}}=\frac{1}{2}\dot{M}_{\mathrm{wind}}v_{\mathrm{rad,max}}^2\approx 4.3\times 10^{43}\ \mathrm{erg\ s}^{-1}$.

\section{Summary}\label{Summary}

J1259+1213 is a rare QSO hosting blueshifted BALs of Fe II, H I Balmer series and metastable He I. We find significant variations of the absorption profiles among the SDSS, BOSS and TripleSpec spectra which were taken over a timescale of 5.8 years in the QSO's frame. In the BOSS spectrum, the Balmer series, the He I* line, and the optical Fe II absorptions, which were first identified in the spectra of QSOs, have been reliably measured and the physical conditions of the absorber can be well-constrained using the photoionization simulations. This allows us to explore the dominant mechanism causing the observed BAL variations. It is demonstrated that the changes in the ionizing flux of the absorber alone cannot reproduce the variability. Conversely, assuming an increase in the covering factor of the absorption at higher outflowing velocity can consistently explain the BAL variations. This leads us to conclude that the BAL variability in J1259+1213 is due to the movement of absorbing gas across LOS to the central source.

This work is supported by the National Basic Research Program of China (the 973 Program 2013CB834905) and the National Natural Science Foundation of China (NSFC-11421303 and 11473025). X. Shu is supported by NSFC-11573001. S. Zhang is supported by NSFC-11573024. This research uses data obtained through the Telescope Access Program (TAP), which has been funded by the Strategic Priority Research Program The Emergence of Cosmological Structures (Grant No. XDB09000000), the National Astronomical Observatories, the Chinese Academy of Sciences, and the Special Fund for Astronomy from the Ministry of Finance.

We specially acknowledge the help from Dr Y. Naz\'{e} and his group to share with us the multi-epoch spectroscopic data for the emission-line star Hen 3-209, which consolidates our identification of the optical Fe II absorption lines.

\clearpage

\begin{table}
\begin{center}
\caption{The atomic data for the optical Fe II absorptions detected in J1259+1213.\label{OptFeII}}
\begin{tabular}{ccccc}
\tableline\tableline
Transition & Rest Wavlength (\AA) & Oscillator Strength & Lower Level & Excited Energy (eV)\\
\tableline
FeII 4233 & 4233.172 & $2.59\times 10^{-3}$ & b$^4$P$_{5/2}$ & 2.583\\
FeII 4549 & 4549.474 & $2.33\times 10^{-3}$ & b$^4$F$_{7/2}$ & 2.828\\
FeII 4584 & 4583.837 & $1.82\times 10^{-3}$ & b$^4$F$_{9/2}$ & 2.807\\
FeII 4924 & 4923.927 & $1.04\times 10^{-2}$ & a$^6$S$_{5/2}$ & 2.891\\
FeII 5018 & 5018.440 & $7.5\times 10^{-3}$ & a$^6$S$_{5/2}$ & 2.891\\
FeII 5169 & 5169.033 & $2.26\times 10^{-2}$ & a$^6$S$_{5/2}$ & 2.891\\
FeII 5276 & 5276.002 & $1.26\times 10^{-3}$ & a$^4$G$_{9/2}$ & 3.199\\
FeII 5317 & 5316.615 & $1.37\times 10^{-3}$ & a$^4$G$_{11/2}$ & 3.153\\
\tableline
\end{tabular}
\end{center}
\end{table}

\clearpage

\clearpage

\clearpage

\clearpage

\clearpage

\end{document}